\documentclass[%
 aip,
 amsmath,amssymb,
 reprint,%
]{revtex4-1}

\usepackage{graphicx}
\usepackage{dcolumn}
\usepackage{bm}

\usepackage[utf8]{inputenc}
\usepackage[T1]{fontenc}
\usepackage{mathptmx}
\usepackage{etoolbox}

\usepackage{textcomp}
\usepackage{multirow}
\usepackage{xfrac}
\usepackage{physics}

\usepackage[version=4]{mhchem}

\makeatletter
\def\@email#1#2{%
 \endgroup
 \patchcmd{\titleblock@produce}
  {\frontmatter@RRAPformat}
  {\frontmatter@RRAPformat{\produce@RRAP{*#1\href{mailto:#2}{#2}}}\frontmatter@RRAPformat}
  {}{}
}%
\makeatother
\begin{document}

\title{Photochemical Anisotropy and Direction-dependent Optical Absorption in Semiconductors}

\author{Chiara Ricca}
\affiliation{%
Department of Chemistry and Biochemistry, University of Bern, Freiestrasse 3, CH-3012 Bern, Switzerland 
}%
\affiliation{%
National Centre for Computational Design and Discovery of Novel Materials (MARVEL), Switzerland
}%

\author{Ulrich Aschauer}
\email{ulrich.aschauer@unibe.ch}
\affiliation{%
Department of Chemistry and Biochemistry, University of Bern, Freiestrasse 3, CH-3012 Bern, Switzerland 
}%
\affiliation{%
National Centre for Computational Design and Discovery of Novel Materials (MARVEL), Switzerland
}%

\date{\today}

\begin{abstract}
Photochemical reactions on semiconductors are anisotropic, since they occur with different rates on surfaces of different orientation. Understanding the origin of this anisotropy is crucial to engineering more efficient photocatalysts. In this work, we use hybrid density functional theory (DFT) to identify the surfaces associated with the largest number of photo-generated carriers in different semiconductors. For each material we create a spherical heat map of the probability of optical transitions at different wave vectors. These maps allow to identify the directions associated with the majority of the photo-generated carriers and can thus be used to make predictions about the most reactive surfaces for photochemical applications. Results indicate that it is generally possible to correlate the heat maps with the anisotropy of the bands observed in conventional band-structure plots, as previously suggested. However, we also demonstrate that conventional bands-structure plots do not always provide all the informations and that taking into account the contribution of all possible transitions weighted by their transition dipole moments is crucial to obtain a complete picture.
\end{abstract}

\maketitle

\section{Introduction\label{sec:intro}}

Since Fujishima and Honda for the first time reported \ce{H2} production from water using \ce{TiO2}-based photoelectrodes in 1972~\cite{fujishima1972electrochemical}, semiconductor-based heterogeneous photocatalysis has received great attention due to the increasing energy demand and diminishing fossil resources. Indeed, heterogeneous photocatalytic processes, such as photocatalytic water splitting (water oxidation and hydrogen reduction), artificial photosynthesis (\ce{CO2} reduction), or the degradation of pollutants, constitute one of the most promising solutions for environmental and energy sustainability via direct harvesting of solar energy~\cite{Xie2016, Li2016, Tee2017, SanMartin2020}.

\begin{figure*}[ht]
	\centering
	\includegraphics[width=\textwidth]{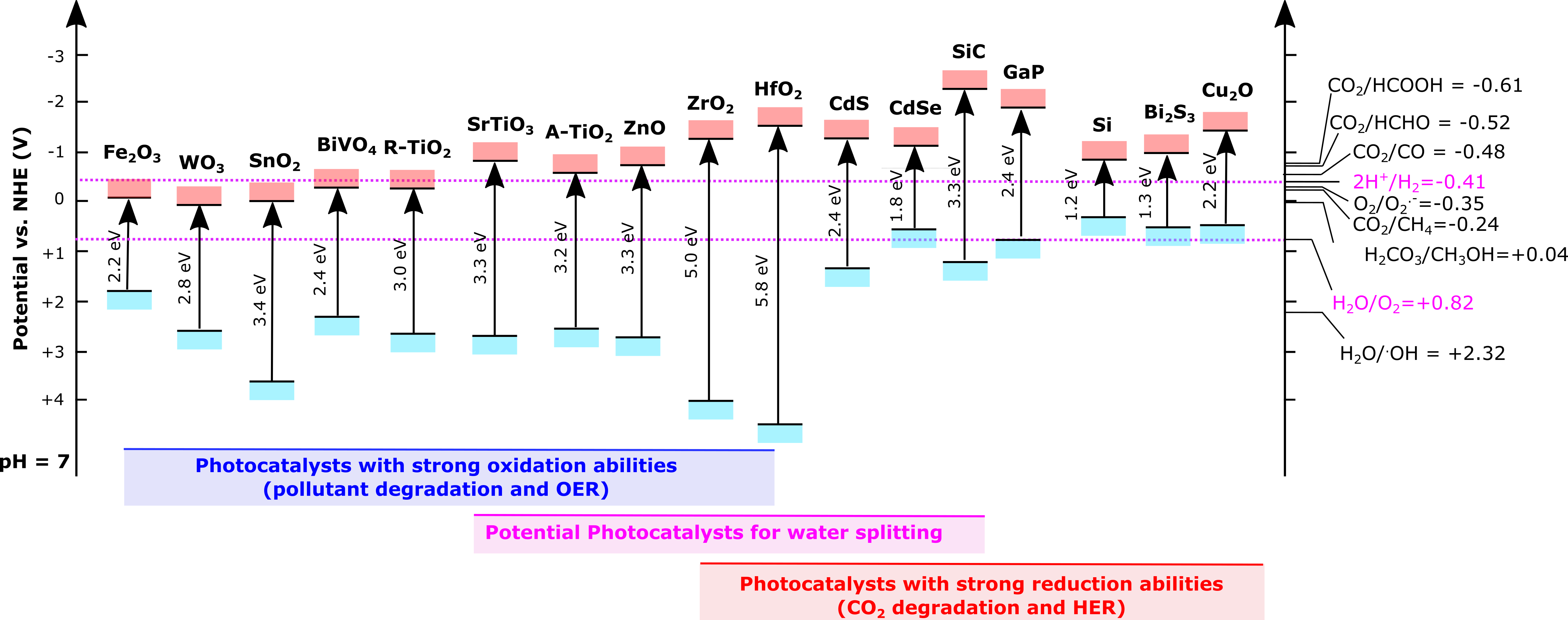}
	\caption{Band-edge positions of some typical semiconductor photocatalysts relative to the energy levels of the redox couples involved several photocatalytic processes. Adapted from Ref.~\onlinecite{Li2016}.}
	\label{fig:photocatalysts}
\end{figure*}

It is well accepted that photocatalytic reactions involve three sequential processes. First, electron/hole pairs are photo-generated when the semiconductor is illuminated by light with an energy equal or greater than the material's bandgap, which leads to excitation of electrons from the valence (VB) to the conduction band (CB). While recombination of electron/hole pairs can occur in the bulk, a fraction of the generated electrons and holes migrate, in a second step, towards the surfaces of the catalyst, where they can, in a final step, initiate reduction and oxidation reactions respectively. However, these reactions can be driven by the photo-generated electrons and holes only if the reduction and oxidation potentials of the targeted reaction lie within the CB and VB of the material, respectively. According to the relative position of the VB and CB and the redox potential of specific reactions, it is possible to identify different types of catalysts: i) strongly oxidative semiconductors with sufficiently positive VB potentials to produce OH radicals such as \ce{SnO2}, \ce{BiVO4}, \ce{TiO2}, \ce{WO3} (see left side of Fig.~\ref{fig:photocatalysts}), that are promising candidates for the photo-degradation of organic pollutants; ii) strongly reductive catalysts with CB position more negative than the hydrogen evolution and \ce{CO2} reduction potentials, such as \ce{Cu2O}, \ce{Bi2Se3}, \ce{SiC}, and \ce{Si} (see right side of Fig.~\ref{fig:photocatalysts}), which are promising semiconductors for solar-fuel production under visible light; and iii) materials with VB/CB potential more positive/negative than the \ce{H2}/\ce{O2} evolution potentials that are suitable to catalyze the overall water splitting reaction (see pink shaded area in Fig.~\ref{fig:photocatalysts})~\cite{Li2016}.

Hundreds of materials are currently available for different photocatalytic applications, however, so far, no semiconductor is able to meet all the requirements in terms of efficiency, stability, safety, cost and ability to absorb solar energy. Efforts to design new materials or to improve the performance of already existing ones can benefit from a deeper understanding of the fundamental properties and mechanisms leading to the observed behavior. For example, it is well known that photochemical reactions on semiconductors are anisotropic, since they occur with different rates on surfaces of different orientation that, hence, have different photocatalytic activities towards reduction and oxidation reactions~\cite{Lowekamp1998, Giocondi2001, Ohno2002, Taguchi2003, Giocondi2007}. Despite this phenomenon being fundamental for the design of photocatalysts, the underlying mechanism is not well understood~\cite{Giocondi2007}. Giocondi \textit{et al.}~\cite{Giocondi2007} suggested, as a general principle, that the observed anisotropic photochemical reactivity can be explained by the spectral distribution of the incident light and the anisotropic dispersion of the photocatalyst's electronic band structure: the lowest energy photons that are able to initiate optical transitions will lead to charge carriers with wave vectors aligned with the transition's position in reciprocal space. The photo-generated carriers will therefore preferentially migrate to surfaces perpendicular to that wave vector, leading to an increased reactivity of these surfaces. Using this concept, they explained the observed larger activity of the (100) surface of \ce{SrTiO3} micro crystals compared to the (110) and (111) surfaces towards both oxidation and reduction reactions. Photons with energies larger than the band-gap (3.25 eV) and up to 4.1 eV (maximum within the experimental set-up) can excite transitions near the $\Gamma$-point, some along the [110] and [111] directions, but the large majority along the [100] direction where the bands are weakly dispersive (flat). The electronic band structure of the material thus favors the creation of photo-generated carriers with wave vector perpendicular to the (100) surface. The authors suggest that this is a fundamental principle applicable not only to \ce{SrTiO3}, but to all photocatalysts. We note here that, as explained in Ref.~\onlinecite{Giocondi2007}, there can be several reason for the surface perpendicular to which most carriers are created to not appear as most reactive in experiments: phonon scattering resulting in a change of the carrier momentum and recombination of the carriers in the bulk or at surface or defect states that act as traps are only some of the possible explanations. Nevertheless, the principle reflects a fundamental property of photocatalysts and can be invoked to predict the photochemical anisotropy of semiconductor materials.

DFT is a valuable tool for material design and discovery. In particular, hybrid DFT can provide accurate structural and electronic properties of semiconductors~\cite{Cora2004, Brothers2008, Henderson2011}. It can also be used to predict optical properties of materials~\cite{Gajdos2006, Wang2019, Nishiwaki2020}. Quantities like the absorption coefficient can be derived from the frequency-dependent dielectric function, computed within the independent particle approximation. This consists of an imaginary part determined as a summation over CB states, and a real part, obtained from the Kramers-Kronig transformation~\cite{Gajdos2006}. According to Fermi's Golden Rule, the optical absorption of a semiconductor at a photon energy $\hbar\omega$ is proportional to~\cite{dresselhaus2001solid}
\begin{equation}
	\frac{2\pi}{\hbar} \frac{1}{4\pi^3} \int \left|\bra{v_{\vec{k}}}\vec{p}\ket{c_{\vec{k}}}\right|^2 \delta(E_{c,\vec{k}} - E_{v,\vec{k}}) -\hbar\omega d\vec{k},
	\label{eq:fermigoldenrule}
\end{equation}
where the Brillouin-zone integral sums the joint density of states at energy $\hbar\omega$ over different $\vec{k}$ vectors and $\bra{v_{\vec{k}}}\vec{p}\ket{c_{\vec{k}}}$ is the transition dipole moment (TDM) matrix, i.e. the electric dipole moment associated with a transition from a valence state $v_{\vec{k}}$ to a conduction state $c_{\vec{k}}$. The direction of TDM gives the polarization of the transition and determines how the system will interact with an electromagnetic wave of a given polarization, while the square of its norm represents the probability for a transition between the two states to occur.

In this work, we use hybrid DFT to investigate the anisotropy of the optical absorption of different photocatalysts by creating spherical heat maps of optical-transition probabilities along different directions in space, when the material interacts with photons within a specific energy range. These direction-dependent maps of the optical absorption identify wave vector associated with the majority of photo-generated carriers upon irradiation with light of a specific energy. Hence, they can be used to make predictions about the most reactive surfaces for photochemical applications, as suggested in Ref.~\onlinecite{Giocondi2007}. However, differently from Ref.~\onlinecite{Giocondi2007}, our predictions do not only reflect the dispersion of the material's electronic band structure as shown in conventional two-dimensional (2D) band structure plots, but they take into account the full three-dimensional (3D) band structure as well as the probability for each transition to occur (transition allowed or forbidden by the selection rules) via the calculation of the TDM. Indeed, while in the majority of cases, results can be explained by a straightforward comparison of the heat map with the 2D electronic band structure, in line with the hypothesis suggested of Giocondi \textit{et al.}, more generally, taking into account all possible transitions weighted by the corresponding transition dipole moments is crucial to obtain a complete picture.

\section{Methods\label{sec:compdetails}}

DFT calculations were carried out with the Vienna Ab-initio Simulation Package (VASP)~\cite{Kresse199347, Kresse1994, KRESSE199615, Kresse199654} using the Heyd-Scuseria-Ernzerhof (HSE)~\cite{HeydJCP2003, HeydJCP2004} hybrid exchange-correlation functional. Projector augmented-wave potentials~\cite{BlochlPRB1994, KressePRB1999} were used with a plane-wave cut-off of 500 eV. Space group, magnetic properties, and dimension of the $\Gamma$-centered $k$-mesh used to sample the Brillouin zone, as well as the type of pseudopotential and the percentage of exact exchange used to model the primitive cell of the various materials are reported in Table~\ref{tbl:compdetails} of the supporting information (SI). All internal and lattice parameters were relaxed with a force threshold of 10$^{-3}$~eV/\AA. This setup provides an excellent agreement with experiment (see SI Section~\ref{sec:MethodSI}). Optical properties of the materials were obtained via the calculation of the frequency-dependent dielectric function within the independent particle approximation~\cite{Gajdo2006} and analyzed using the pymatgen Wavederf module~\cite{Ong2013}. 

The direction-dependent probability of a material to absorb photons in a given wavelength range ($\lambda_{min}$ to $\lambda_{max}$) was mapped on spherical heat maps, obtained by projecting the k-mesh used to sample reciprocal space on the surface of a unit sphere centered at the origin of the reciprocal lattice. Each k-point is associated with an absorption coefficient ($P_k$) computed by summing over the square of the TDM for transitions between bands $v_{\vec{k}}$ and $c_{\vec{k}}$ at that k-point $\vec{k}$, if the wavelength ($\lambda_{vc}$) of the transition is within the considered energy range ($\lambda_{min}$ to $\lambda_{max}$):
\begin{equation}
	P_{\vec{k}}= \sum_{vc}^{\lambda_{min} < \lambda_{vc} < \lambda_{max}} |\bra{v_{\vec{k}}}\vec{p}\ket{c_{\vec{k}}}|^2.
\label{eq:heatmap}
\end{equation}
The final spherical heat map of direction-dependent absorption is created by interpolating the obtained discrete data onto a regular grid in spherical coordinates. We considered transitions in the energy range going from the bandgap ($E_\textrm{g}$) of each material up to a minimum energy ($E_\textrm{g}+\Delta E$) necessary to observe vertical transitions (cf. Table ~\ref{tbl:compdetails}).

\section{Results and Discussion\label{sec:results}}
%
\begin{figure}
	\centering
	\includegraphics[width=\columnwidth]{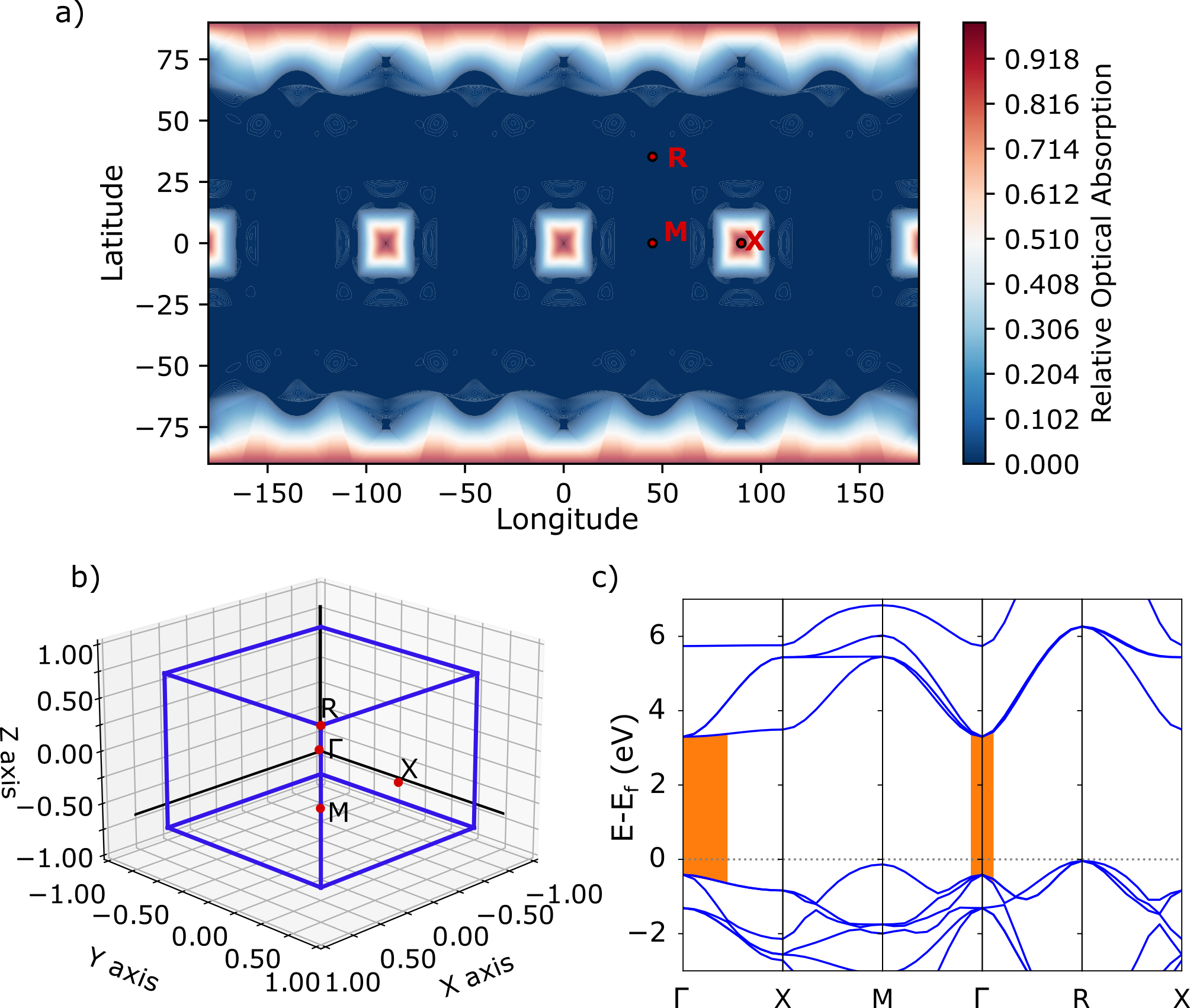}
	\caption{a) Heat map of the direction-dependent optical absorption shown as a cylindrical projection along with the position and labels of the high symmetry points, b) Brillouin zone and c) band-structure plot for \ce{SrTiO3}. The shaded regions indicate momentum states where vertical transitions are possible when the photon energy is equal to $E_g$(\ce{SrTiO3})+0.8~eV.}
	\label{fig:STO}
\end{figure}
Giocondi \textit{et al.}~\cite{Giocondi2007} investigated the anisotropic reactivity of \ce{SrTiO3} using photochemical reactions that deposit insoluble products on the surface of \ce{SrTiO3} micro crystals. Their results indicated that $\lbrace 100 \rbrace$ surfaces are most reactive. This observation was explained considering that the majority of electron-hole pairs used for photochemical reactions are created by photons with energies greater than the band gap but lower than 4.1~eV and that, according to the band structure of \ce{SrTiO3}, photons with these energies can excite a larger number of transitions along $[100]$ than other directions, due to the weakly dispersive bands along this direction.

The heat map of Fig.~\ref{fig:STO}a) shows the direction-dependent probability of \ce{SrTiO3} to absorb photons with energies up to 4.1~eV, computed according to the procedure described in Sec.~\ref{sec:compdetails}. In agreement with the experimental observations, we predict the $\lbrace 100 \rbrace$ surfaces to be most active for photochemical processes, since the largest absorption is observed projected on this direction. As suggested in Ref.~\onlinecite{Giocondi2007}, this result can be rationalized using arguments based on the computed band structure: photons with energies between the \ce{SrTiO3} bandgap and 4.1~eV can excite transitions near the $\Gamma$-point, with excited states in the $[100]$ ($\Gamma$ to X, cf. Fig.~\ref{fig:STO}b) and Fig.~\ref{fig:BS} in the SI), $[110]$ ($\Gamma$ to M), or $[111]$ ($\Gamma$ to R) directions as shown by the shaded regions in Fig.~\ref{fig:STO}c). However, the flat character of the bands in the $\Gamma$ to X direction results in a much larger number of photo-generated carriers with wave-vector perpendicular to the (100) surface, explaining the larger absorption intensity along this direction.

\begin{figure}
	\centering
	\includegraphics[width=\columnwidth]{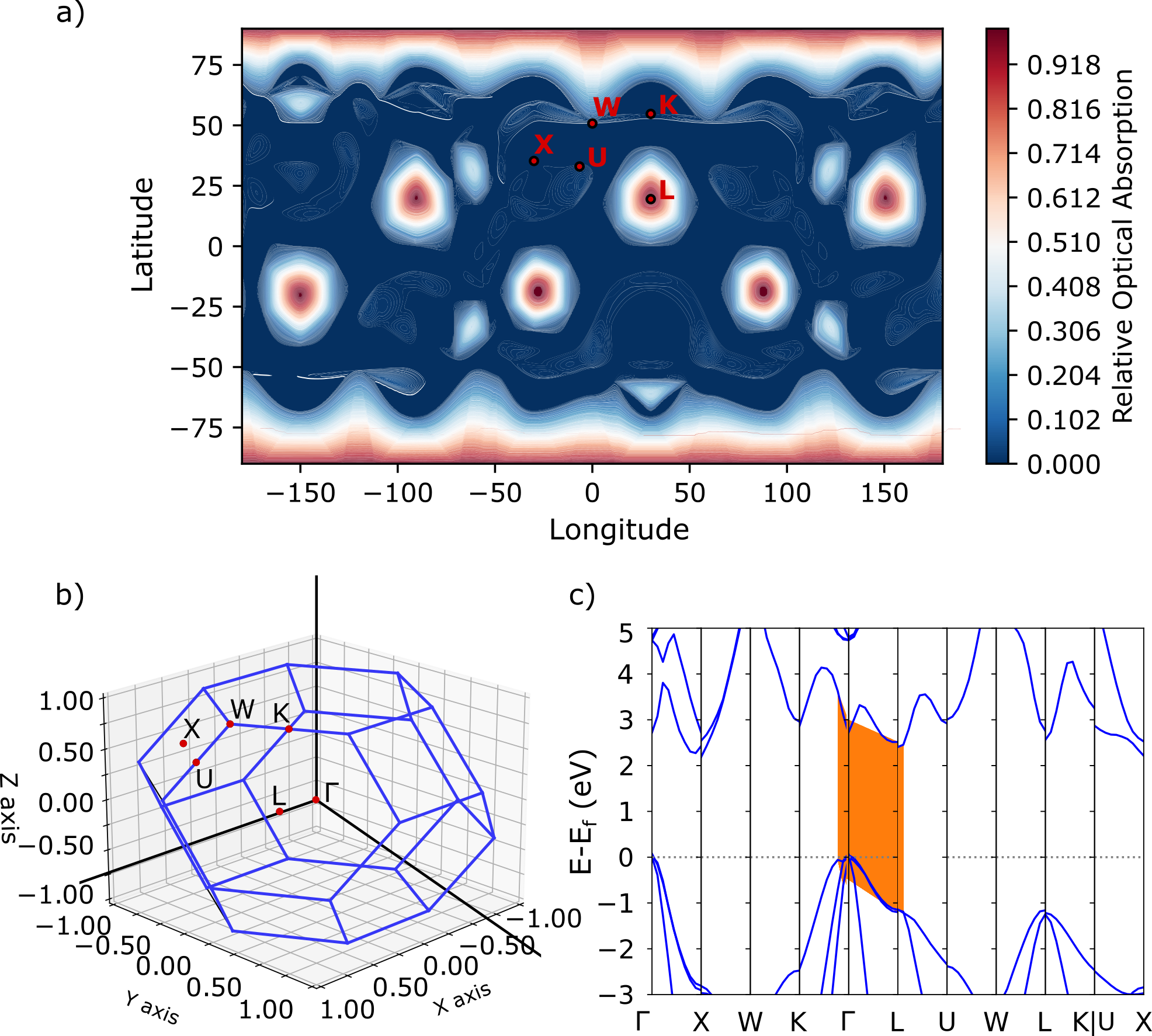}
	\caption{a) Spherical heat map of the direction-dependent optical absorption shown as a cylindrical projection along with the position and labels of the high symmetry points, b) Brillouin zone and c) band-structure plot for GaP. The shaded regions indicate momentum states where vertical transitions are possible when the photon energy is equal to $E_g$(\ce{GaP})+1.5~eV.}
	\label{fig:GaP}
\end{figure}

When the same procedure is applied to other photocatalyst materials, the argument based on the conventional 2D band-structure plots is verified in the majority of cases. For example, our heat map for GaP in Fig.~\ref{fig:GaP}a confirms the suggestion by Giocondi \textit{et al.}~\cite{Giocondi2007} that for materials with zinc-blende structure the most reactive orientation should be (111). Indeed, photons with energies between the GaP bandgap and 4.0~eV can excite transitions near the $\Gamma$-point with excited states in the $[111]$ direction ($\Gamma$ to L, cf. Fig.~\ref{fig:GaP}c) and Fig.~\ref{fig:BS} in the SI).

\begin{figure}
	\centering
	\includegraphics[width=\columnwidth]{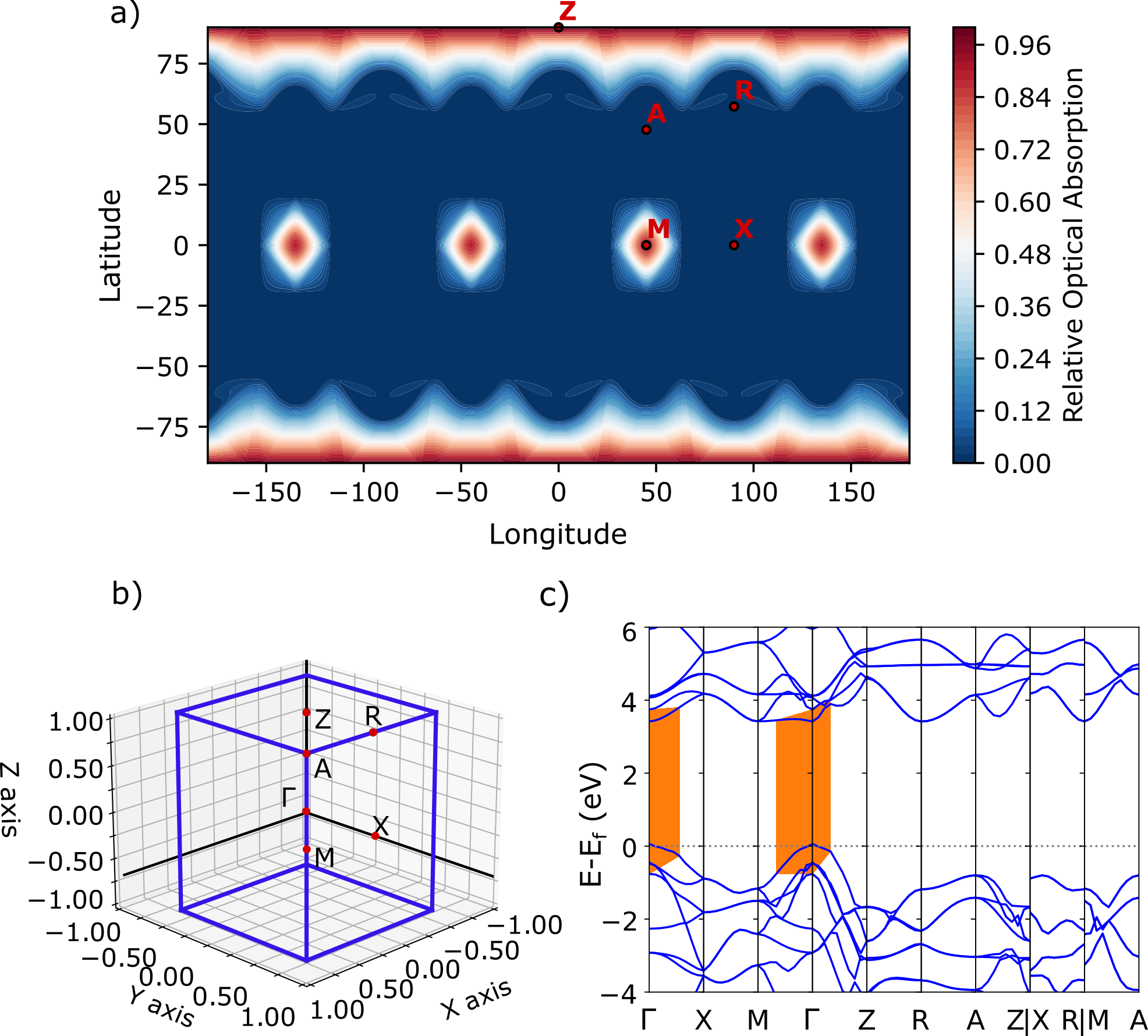}
	\caption{a) Spherical heat map of the direction-dependent optical absorption shown as cylindrical projection along with the position and labels of the high symmetry points, b) Brillouin zone and c) band-structure plot for \ce{TiO2} rutile. The shaded regions indicate momentum states where vertical transitions are possible when the photon energy is equal to $E_g$(\ce{TiO2}-rutile)+0.8~eV.}
	\label{fig:TiO2R}
\end{figure}

However, there are materials for which the conventional 2D band structure does not seem to provide all the necessary information to correctly predict the most reactive surface orientation. For example, the band structure computed for the rutile polymorph of \ce{TiO2} in Fig.~\ref{fig:TiO2R}c) suggests that the largest number of excited transitions induced by photons of energy larger than the bandgap (E$_\textrm{g}$) up to E$_\textrm{g}$+0.8~eV takes place from $\Gamma$ to X ($[100]$ direction) and from $\Gamma$ to M ($[110]$ direction) and with a smaller contribution also from $\Gamma$ to Z ($[001]$ direction). Do due the less dispersive character of the 2D bands along the first two directions, one would expect the (110) and (100) surfaces to be more reactive than the (001) surface. The heat map of the direction-dependent optical absorption of Fig.~\ref{fig:TiO2R}a) suggests, instead, that photo-generated carriers with wave-vectors perpendicular to this latter surface should be most abundant, followed by carriers with wave-vectors perpendicular to $[110]$. In order to explain this apparent contradiction, we computed the TDM for a vertical transition from the top of the valence band to the bottom of the conduction band at the k-points Z (12.3~Debye$^2$), M (2.0~Debye$^2$), and X (0.2~Debye$^2$) using the VASPKIT package~\cite{VASPKIT}. Based on these TDM values, we conclude that, even though the flattest path in the band structure is from $\Gamma$ to M or $\Gamma$ to X, the larger probability for transitions at the Z point leads to photo-generated carriers with wave-vectors perpendicular to $[001]$ to be most abundant.

\begin{figure}
	\centering
	\includegraphics[width=\columnwidth]{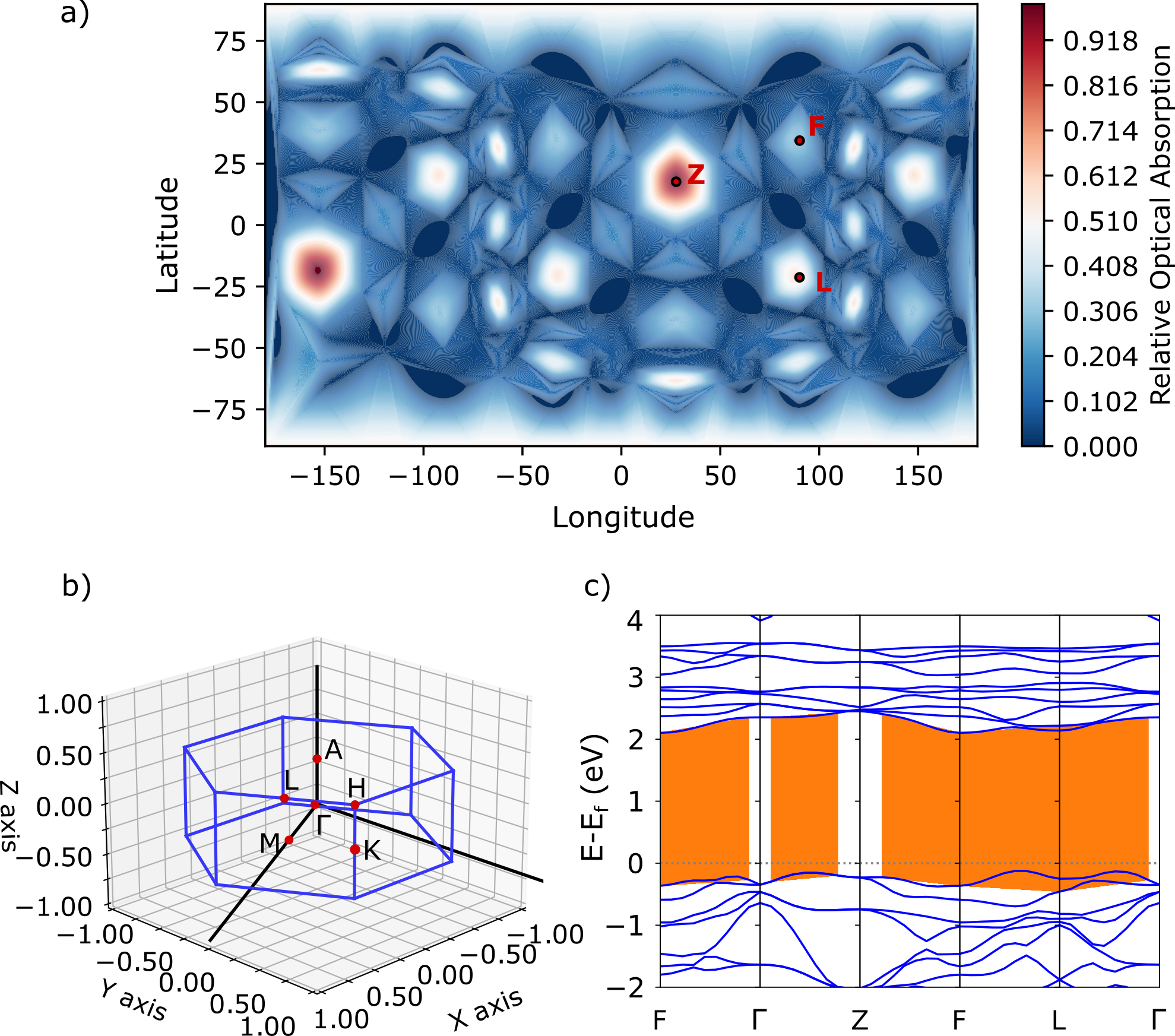}
	\caption{a) Spherical heat map of the direction-dependent optical absorption shown as cylindrical projection along with the position and labels of the high symmetry points, b) Brillouin zone and c) band-structure plot for \ce{Fe2O3}. The shaded regions indicate momentum states where vertical transitions are possible when the photon energy is equal to $E_g$(\ce{Fe2O3})+0.5~eV.}
	\label{fig:Fe2O3}
\end{figure}
These result shed light on the experimental results for rutile that are often contradictory and obtained under different experimental conditions, for different photochemical reactions, and on different types of samples. For example, Ohno \textit{et al.}~\cite{Ohno2002} reported that Pt deposits preferentially on rutile $\lbrace 110 \rbrace$ facets, consistent with the band-structure prediction. Instead, Guennemann \textit{et al.}~\cite{Guennemann2019} showed (001) to be one of the rutile surfaces with the largest photonic efficiency and to be the most reactive surface for methanol oxidation. This is in agreement with our prediction and also with the results of Ahmad \textit{et al.}~\cite{Ahmed2011} reporting the (001) surface to show a higher photocatalytic activity than both the (110) and (100) surfaces for both methanol oxidation and terephthalic acid hydroxylation. Also Lowekamp \textit{et al.}~\cite{Lowekamp1998} reported that $\lbrace 001 \rbrace$ facets show a higher reactivity for \ce{Ag+} reduction  compared to the $\lbrace 110 \rbrace$ and the $\lbrace 100 \rbrace$ facets, even though the $\lbrace 101 \rbrace$ facets are the most active. Larger photocatalytic activity of $\lbrace 101 \rbrace$ surfaces was reported also by Giocondi \textit{et al.}~\cite{Giocondi2001}, Ohno \textit{et al.}~\cite{Ohno2002}, Luttrell \textit{et al.}~\cite{Luttrell2014} for decomposition of methyl orange, or by Guennemann \textit{et al.}~\cite{Guennemann2019} for terephthalic acid hydroxylation. On the other hand, Quah \textit{et al.}~\cite{Quah2010} reported that there is no special photocatalytic activity of this latter facet compared to other rutile surfaces, in line with our heat map that does not suggest the creation of carriers with momentum along this direction.

A similar situation is observed also for \ce{Fe2O3}. While the band structure in Fig.~\ref{fig:Fe2O3}c) shows fairly non-dispersive bands and hence possible transitions with momenta along different directions in space, the heat map of the direction-dependent optical absorption in Fig.~\ref{fig:Fe2O3}a) indicates that the (111) facet (transitions from from $\Gamma$ to Z) should be much more active than other surfaces. This prediction can also be rationalized by the larger TDM for a transition from the top of the VB to the bottom of the CB at the Z point (68.3 Debye$^2$), compared to the other high symmetry k-points, such as L (43.3~Debye$^2$) or F (10.3~Debye$^2$).

\begin{figure}
	\centering
	\includegraphics[width=\columnwidth]{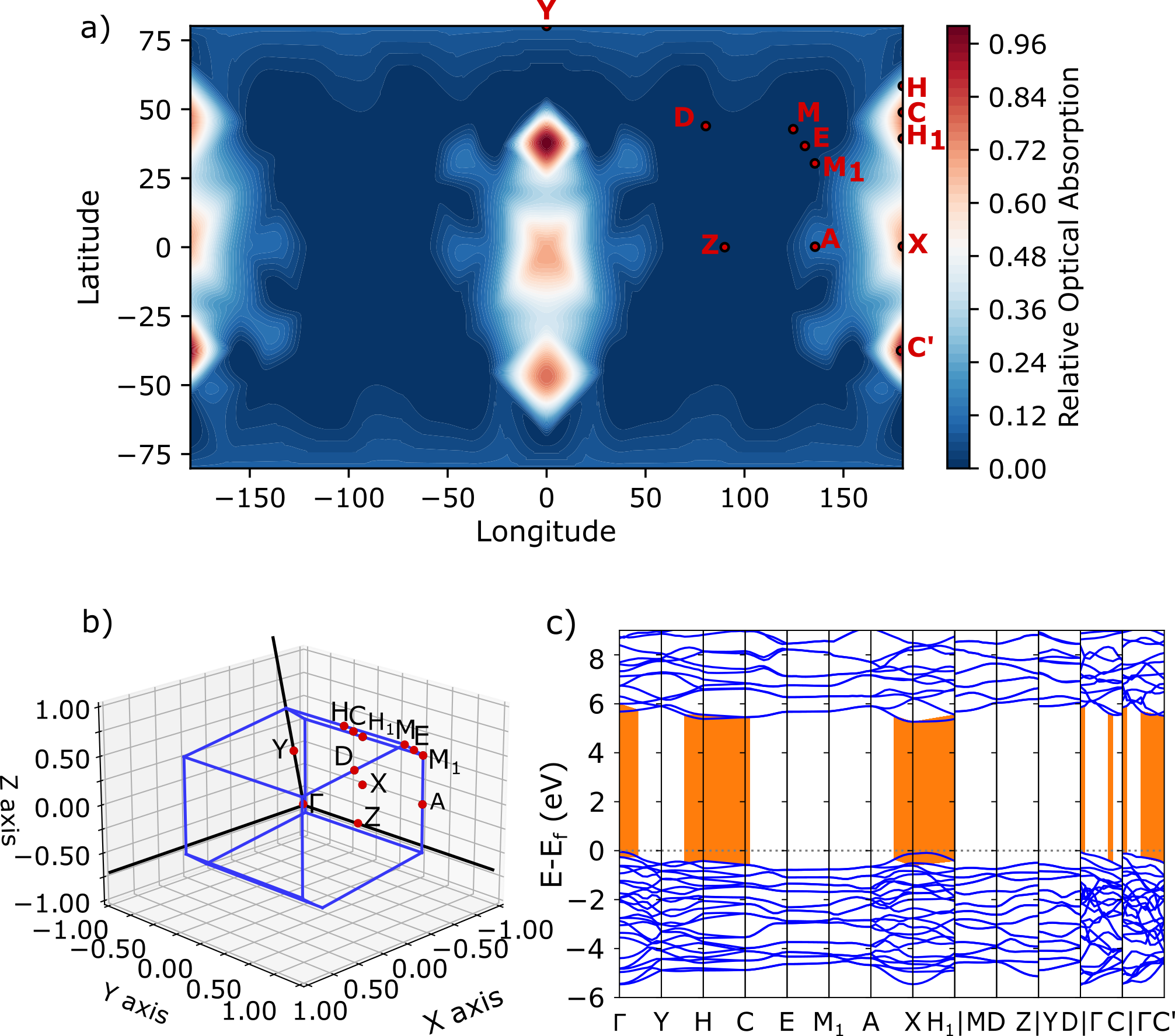}
	\caption{a) Spherical heat map of the direction-dependent optical absorption shown as cylindrical projection along with the position and labels of the high symmetry points, b) Brillouin zone and c) band-structure plot for \ce{ZrO2}. The shaded regions indicate momentum states where vertical transitions are possible when the photon energy is equal to $E_g$(\ce{ZrO2})+0.8~eV.}
	\label{fig:ZrO2}
\end{figure}
Finally, one has to consider that conventional band structures along a high-symmetry path do not show the band dispersion along all directions. For example, the heat map for \ce{ZrO2} in Fig.~\ref{fig:ZrO2}a) suggests that the strongest absorption should occur at a point we indicated as C' ($[\bar{1}0\bar{1}]$ direction). The segment from $\Gamma$ to C' (see Fig.~\ref{fig:ZrO2}c)) is indeed quite non-dispersive and definitely flatter than the segment from $\Gamma$ to C ($[\bar{1}01]$ direction), which is the one with the second strongest optical absorption in Fig.~\ref{fig:ZrO2}a), but which is not reported in conventional band-structure plots for \ce{ZrO2}.

These results not only show that our approach intrinsically takes into account the probability of a transition to happen via the calculation of the TDM, which was not considered in the original approach of Ref.~\onlinecite{Giocondi2007}, but also highlight that using only the dispersion of the conventional band structure to make predictions may not result in a complete description of the direction-dependent absorption properties.
\begin{figure*}[p!]
	\centering
	\includegraphics[width=0.9\textwidth]{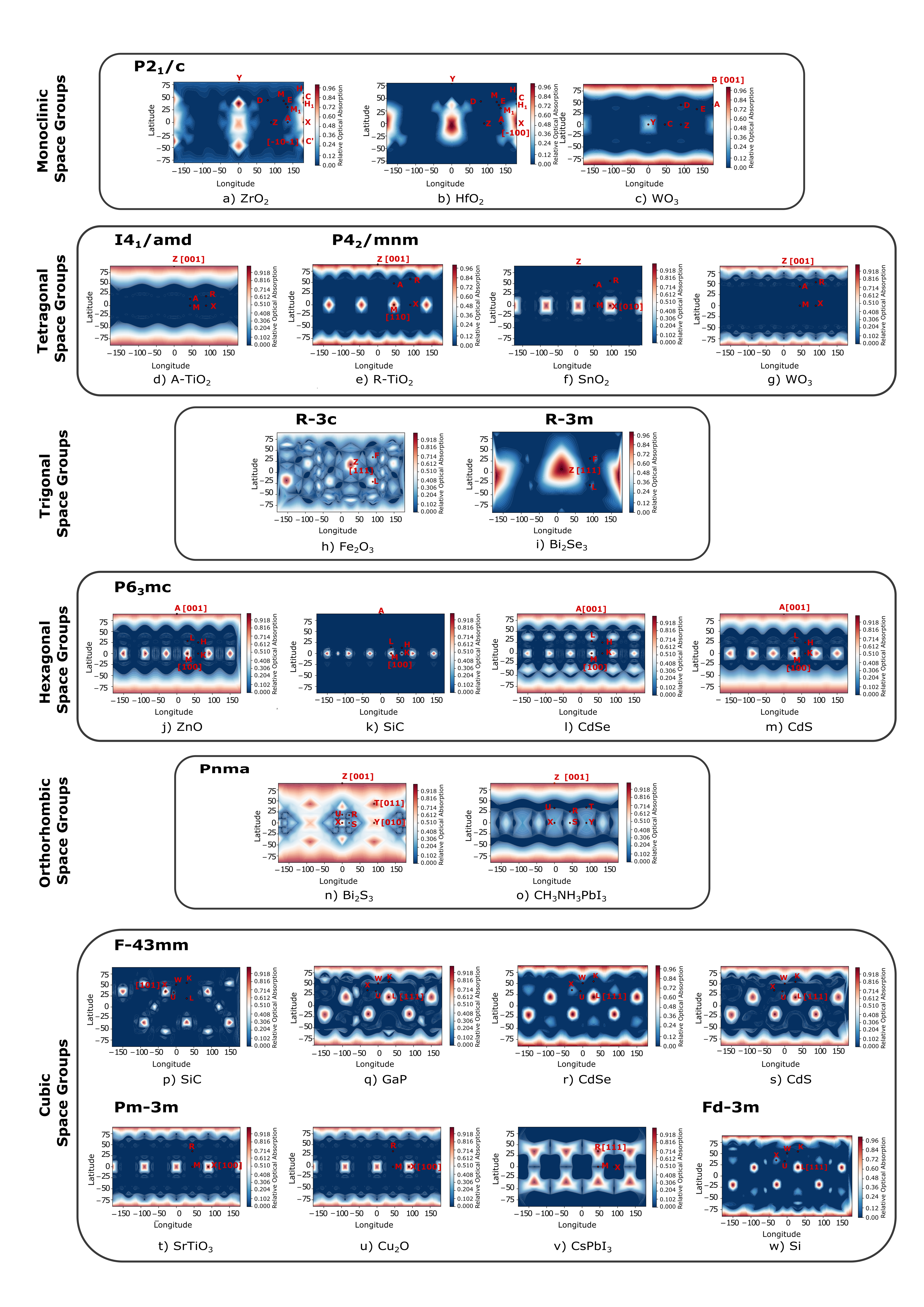}
	\caption{Heat maps of the direction-dependent optical absorption for the selected photocatalysts. Transitions in the energy range from the bandgap ($E_g$) of each material up to E$_g$+$\Delta$E~eV (see Sec.~\ref{sec:compdetails}) and Table S1 in the SI) are taken into account.}
	\label{fig:DDAO}
\end{figure*}
Fig.~\ref{fig:DDAO} reports the heat maps for the selected photocatalysts computed for transitions in the energy range going from the bandgap of each material up to the minimum energy necessary to observe vertical transitions. These heat maps can be used to predict the anisotropic reactivity of different photocatalysts. For example, the (100) surface is predicted to be the most reactive surface for \ce{SrTiO3}, \ce{Cu2O}, \ce{HfO2}, \ce{SnO2}, and the hexagonal \ce{SiC} or together with the (001) surface for ZnO, and the hexagonal CdSe and CdS structures. Strong reactivity of the $\lbrace 001 \rbrace$ facets is predicted for materials like \ce{WO3}, anatase, \ce{CH3NH3PbI3}, and \ce{Bi2S3}. Finally, the (111) surface should be the most reactive not only for materials with the zinc-blend structure, as discussed above, but also for Si, \ce{Bi2Se3}, \ce{Fe2O3}, and \ce{CsPbI3}, while for the cubic SiC phase a larger reactivity is expected for the $\lbrace 001 \rbrace$ facets.

We stress here that our heat maps suggest which surfaces are perpendicular to directions with the largest number of photo-generated carriers, but nothing ensures that these surfaces will necessarily be the most reactive in actual experiments. Phonon scattering can, for example, result in a change of the carrier momentum, implying that carrier mobility and mean free path need to be large for the carrier to reach a surface perpendicular to the initial direction~\cite{Giocondi2007}. 

Furthermore, our approach does not take into account surface band bending and hence the band-edge position at different surfaces nor the possibility of surface or surface defect states to act as trapping or recombination centers~\cite{Giocondi2007}. Once at the surface, photo-generated carriers have to be efficiently transferred to acceptor/donor molecules, which may be preferentially absorbed on a surface different from the one with the largest number of carriers~\cite{Giocondi2007, Tan2016}. The stability and structural properties of the surface are also important; in some cases, the reactivity of a surface can by related to the type and number of exposed undercoordinated surface sites~\cite{Su2017} or the most reactive surface may not be the most stable one~\cite{chang2014functional, chang2014activity, danilovic2014activity, roy2018trends}. For example, the larger photocatalytic activity of anatase nanoparticles with respect to rutile powders has sometimes been explained by the more active anatase (101) and (100) surfaces to be among the thermodynamically most stable surfaces~\cite{Ahmed2011}.

 The mechanism of the studied redox process is also important. The above-mentioned contradictory results for rutile's reactivity were partially explained by some authors investigating oxidation and others reduction reactions. In addition, depending on experimental conditions, a reaction can happen following different mechanisms: for example, methanol is oxidized directly by photo-generated holes for high methanol concentration, while for larger water content, the oxidation is mediated by hydroxyl radicals, thus resulting in different reactivity of rutile surfaces for the same reaction but under different experimental conditions~\cite{Guennemann2019}. 
 
 Finally, the activity of a photocatalyst is strongly dependent on its physicochemical properties, like the particle shape and size~\cite{Tan2016}. Giocondi \textit{et al}.~\cite{Giocondi2007} suggested that the different conclusions of various experimental works on rutile can also be rationalized by having been performed using either extended planar samples or single crystal particles. The reactivity of these two types of samples is hard to compare. A planar sample may exhibit a reduced reactivity compared to a surface with the same orientation on a micro crystal bounded by several different facets, because the planar specimen only contains a subset of the particle's surface sites and may therefore lack highly active sites for a given redox reaction.
 
Despite these limitations, our spherical heat maps contain information about a fundamental material property - the optical absorption and hence ability to photo-generate carriers along a given direction - and can thus be used as a guide in material engineering to enhance the photocatalytic performance of a catalyst.

\section{Conclusions\label{sec:concls}}
In the present work, the direction-dependent optical properties of different photocatalysts were investigated via hybrid DFT calculations to elucidate their anisotropic reactivity. The probability of a crystal to photo-generate carriers with momentum along a certain direction in space upon irradiation with light of a specific energy was mapped on the surface of a unit sphere centered on the origin of the reciprocal lattice to obtain 3D heat maps. For each point of the map, the contribution of all possible excited transitions weighted by their probability (transition-dipole matrix) was taken into account. 

Results indicate that, in the majority of the cases, the pattern of the heat map, and hence the anisotropy of the photo-generated carriers, can be directly explained using the anisotropy of the band dispersion observed in conventional 2D band-structure plots: the flatter the bands along one direction, the larger the number of photo-generated carriers with momentum in that direction, as suggested by Giocondi \textit{et al.}~\cite{Giocondi2007}. However, the results also suggest that 2D band-structure plots are not always sufficient to understand the direction-dependence of the optical properties, but that weighing by a transition's probability according to the the transition-dipole matrix can be crucial to obtain a complete picture.

Despite the possibility that the surface associated with the largest number of photo-generated carriers will not appear as most reactive in experiments, our approach captures a fundamental material property and can thus be invoked to predict the anisotropic reactivity of crystals to guide the design and development of highly efficient photocatalysts.

\begin{acknowledgments}
This research was supported by the NCCR MARVEL, funded by the Swiss National Science Foundation. Computational resources were provided by the University of Bern (on the HPC cluster UBELIX, http://www.id.unibe.ch/hpc) and by the Swiss National Supercomputing Center (CSCS) under projects ID mr26 and s1033.
\end{acknowledgments}

\section*{Data Availability Statement}
The data used to derive the findings in this study are openly available on Materials Cloud at http://doi.org/[doi].

\nocite{*}
\bibliography{references}

\clearpage
\renewcommand{\thetable}{S\arabic{table}} 
\setcounter{table}{0}
\renewcommand{\thefigure}{S\arabic{figure}}
\setcounter{figure}{0}
\renewcommand{\thesection}{S\arabic{section}}
\setcounter{section}{0}
\renewcommand{\theequation}{S\arabic{equation}}
\setcounter{equation}{0}
\onecolumngrid

\begin{center}
\textbf{Supplementary information for\\\vspace{0.5 cm}
\large Photochemical Anisotropy and Direction-Dependent Optical Absorption Properties in Semiconductors\\\vspace{0.3 cm}}
Chiara Ricca,$^{1, 2}$ and Ulrich Aschauer$^{1, 2}$

\small
$^1$\textit{Department of Chemistry and Biochemistry, University of Bern, Freiestrasse 3, CH-3012 Bern, Switzerland}

$^2$\textit{National Centre for Computational Design and Discovery of Novel Materials (MARVEL), Switzerland}

(Dated: \today)
\end{center}

\section{Computational Details\label{sec:MethodSI}}

\begin{figure*}[ht]
	\centering
	\includegraphics[width=0.8\textwidth]{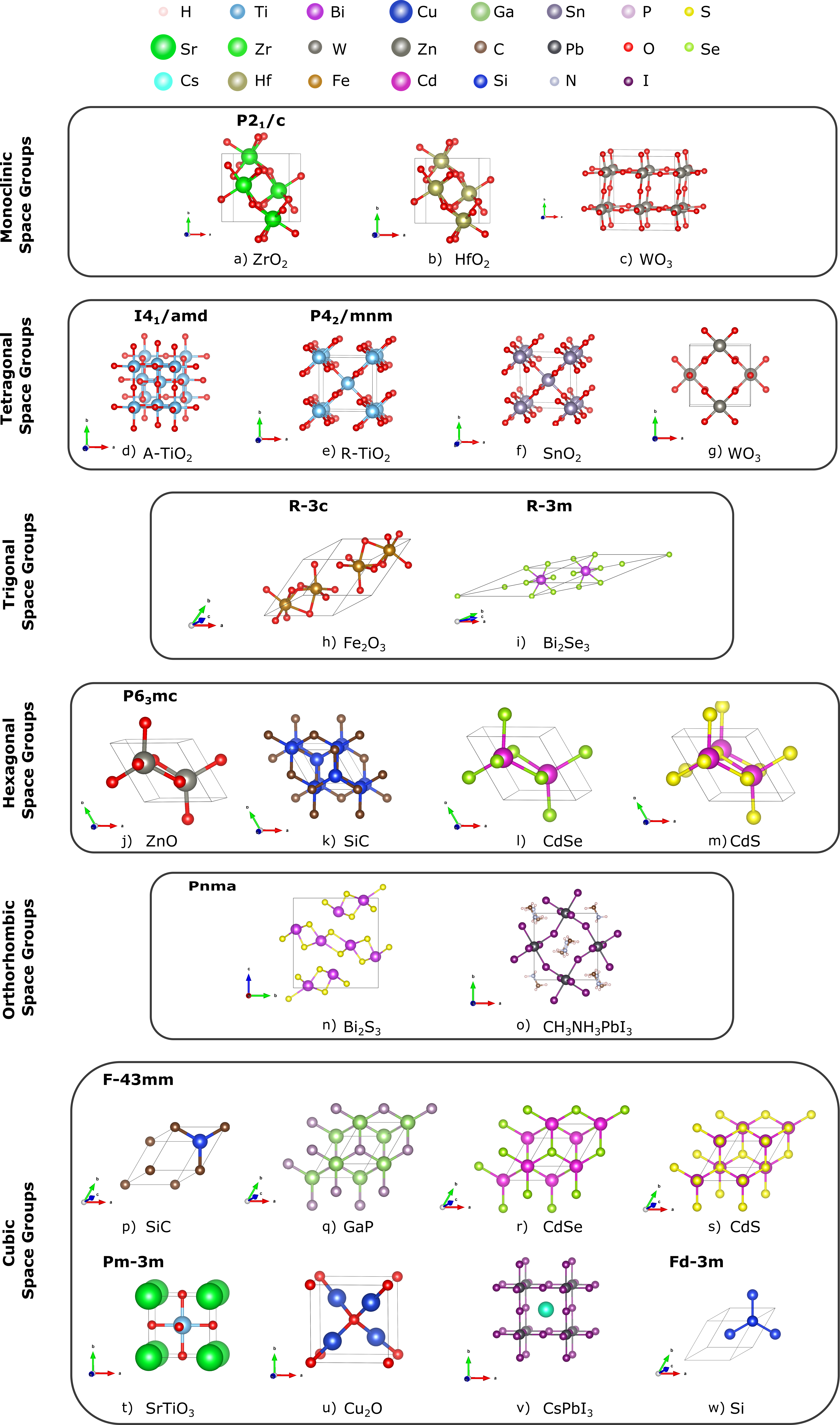}
	\caption{Schematic representation of the primitive cell of the different materials considered in this work.}
	\label{fig:cells}
\end{figure*}

Table \ref{tbl:compdetails} report computational parameters (magnetic order, pseudopotential, k-point mesh, fraction of exact exchange and energy range for transitions), while Fig. \ref{fig:cells} shows the unit cells used for the various materials.

\begin{table}[ht]
\caption{Space group, magnetism, VASP pseudopotentials (Pseudop), dimension of the k-mesh grid (K-points),  percentage of exact exchange (HF\%, in \%) used in the HSE functional, and energy range (\textbf{$\Delta$E (eV)}) used to compute the absorption properties (see main text) for  the studied materials. NM and AFM stand for non magnetic and anti-ferromagnetic, respectively. For \ce{Fe2O3}, the  $+$ and $-$ symbols designate up- and down-spin directions with respect to the $z$-axis of the rhombohedral cell\cite{Sandratskii1996}.  \ce{TiO2}-A and \ce{TiO2}-R indicate the anatase and rutile polytypes.
}
\begin{tabular}{cccccccc}
\hline
\hline
\multicolumn{1}{l}{\textbf{Space   group Type}} & \multicolumn{1}{c}{\textbf{Material}} & \multicolumn{1}{c}{\textbf{Space group}} & \multicolumn{1}{c}{\textbf{Magnetism}} & \multicolumn{1}{c}{\textbf{Pseudop}} & \multicolumn{1}{c}{\textbf{K-points}} & \multicolumn{1}{c}{\textbf{HF\%}} & \multicolumn{1}{c}{\textbf{$\Delta$E (eV)}} \\
\hline
\hline
                                                & \ce{ZrO2}                                  & $P2_1/c $                                   & NM                                     & Zr$\_sv$; O                            & 4x4x4                                 & 25.0       & 0.8                       \\
    & \ce{HfO2}                                  & $P2_1/c $                                    & NM                                     & Hf$\_pv$; O                            & 4x4x4                                 & 25.0                             & 0.75 \\
 \multirow{-4}{*}{Monoclinic}                                               & {\ce{WO3}}                                  & $P2_1/c $                                   & NM                                     & W$\_pv$; O                             & 4x4x4                                 & 25.0         & 0.75                     \\
\hline
                                                & \ce{TiO2}-A                                & $I4_1/amd$                                  & NM                                     & Ti$\_sv$; O                            & 6x6x4                                 & 25.0                   & 0.8           \\
                                                & \ce{TiO2}-R                                & $P4_2/mnm $                                 & NM                                     & Ti$\_sv$; O                            & 4x4x6                                 & 25.0             & 0.8                 \\
                                                & \ce{SnO2}                                  & $P4_2/mnm$                                  & NM                                     & Sn$\_d$; O                             & 6x6x8                                 & 32.0\cite{Yang2016}  & 1.0\\

\multirow{-5}{*}{Tetragonal}                    & \ce{\ce{WO3}}                                   &$ P4_2/mnm  $                                & NM                                     & W$\_pv$; O                             & 6x6x8                                 & 25.0           & 0.8                  \\
\hline
                                                & \ce{Fe2O3}                                 & $R\bar{3c}$                                     & AFM (+ -- -- +)                               & Fe; O                                & 6x6x6          & 15.0\cite{ZHOU2016}            &0.5                  \\

\multirow{-2}{*}{Trigonal}                      & \ce{Bi2Se3}                                & $R\bar{3m}$                                   & NM                                     & Bi; Se                               &  4x4x4          & 25.0          &1.0                    \\
\hline
                                                & ZnO                                   & $P6_3mc$                                    & NM                                     & Zn;O                                 & 8x8x4                                 & 37.5\cite{Bashyal2018}          & 1.5                   \\
                                                & SiC                                   & $P6_3mc$                                    & NM                                     & Si; C                                & 8x8x2                                 & 25.0    &2.00                          \\
                                                & CdS                                   & $P6_3mc$                                   & NM                                     & Cd; S                                & 6x6x4                                 & 25.0       & 1.5                       \\

\multirow{-4}{*}{Hexagonal}                     & CdSe                                  & $P6_3mc$                                   & NM                                     & Cd; Se                               & 6x6x4                                 & 25.0      & 1.5                        \\
\hline
                                                & \ce{Bi2S3}                                 & $Pnma$                                     & NM                                     & Bi; Se                               & 6x2x2                                 & 25.0            & 0.8                  \\
\multirow{-2}{*}{Orthorhombic}                                               & \ce{CH3NH3PbI3}                            & $Pnma$                                     & NM                                     & C; H; N;Ob$\_d$;I                      &4x4x2          & 25.0        & 0.8                       \\    
\hline               
                                                & Si                                    & $Fd\bar{3}m $                                   & NM                                     & Si                                   & 8x8x8                                 & 25.0      & 2.5                        \\
                                                & GaP                                   & $F\bar{4}3m$                                    & NM                                     & Ga$\_d$; P                             & 6x6x6                                 & 25.0      &1.5                        \\
                                                & CdS                                   & $F\bar{4}3m$                                    & NM                                     & Cd; S                                & 6x6x6                                 & 25.0     & 1.5                         \\
                                                & CdSe                                  & $F\bar{4}3m$                                   & NM                                     & Cd; S                                & 6x6x6                                 & 25.0    &2.5                          \\
                                                & SiC                                   & $F\bar{4}3m$                                   & NM                                     & Si; C                                & 8x8x8                                 & 25.0          & 4.0                    \\
                                                & \ce{Cu2O}                                  & $Pm\bar{3}m$                                   & NM                                     & Cu; O                                & 6x6x6                                 & 25.0        & 0.8                      \\
                                                & \ce{\ce{SrTiO3}}                                & $Pm\bar{3}m$                                    & NM                                     & Sr$\_sv$; Ti$\_sv$;O                     & 8x8x8                                 & 25.0     & 0.8                         \\
\multirow{-8}{*}{Cubic}                         & \ce{CsPbI3}                                & $Pm\bar{3}m$                                    & NM                                     & Cs$\_sv$; Pb$\_d$; I                     & 4x4x4                                 & 25.0  & 0.8   \\
\hline 
\hline                       
\end{tabular}
\label{tbl:compdetails}
\end{table}

\clearpage

\clearpage
\section{Structural and Electronic Properties of the Photocatalysts \label{sec:MBulkpropertiesSI}}
Table~\ref{tbl:structure} shows the comparison of the experimental lattice parameters with the ones computed using the setup reported in Table~\ref{sec:MethodSI} for the selected photocatalysts. Results are in excellent agreement with experiments, the maximum relative error being lower than 3\% and with a mean absolute error (MAE) of 0.07~\AA, close to the one of 0.04~\AA\ reported in Ref~\onlinecite{Heyd2005}, where the performance of the HSE functional has been assessed using a set of 40 simple and binary semiconductors.

Hybrid functionals are often used to accurately describe the electronic properties of semiconductors, because  they incorporate a fraction of exact Fock exchange into standard DFT functionals, allowing to correct for  self-interaction errors.  However, their application is  still quite expensive, especially for large cells and when plane-wave pseudopotentials are used as basis set. Table~\ref{tbl:bandgap} suggests a very good agreement between the computed and experimental bandgaps with a mean absolute error of 0.17~eV, in good agreement with the MAE of 0.26~eV reported in  Ref~\onlinecite{Heyd2005}. The largest error are obtained for materials containing heavy elements (Bi, I, or Pb), for which we observed the  well-known tendency of HSE to overestimate their bandgap  due to spin-orbit coupling effects~\cite{Savory2016}. 

\begin{table}[ht]
\caption{Comparison of the computed ($a$, $b$, and $c$, in  \AA)  and experimental ($a_{expt}$, $b_{expt}$, and $c_{expt}$, in \AA) lattice parameters. The subscript  $c$, $h$, $o$, $t$, $z$ stands for cubic, hexagonal, orthorhombic, tetragonal, and zinc-blend-type polymorphs, respectively. \ce{TiO2}-A and \ce{TiO2}-R  indicate the anatase and rutile polytypes.
}
\begin{tabular}{lcccccccl}
\hline 
\hline 
\textbf{Material} & \textbf{Space Group} & $a$ & $a_{expt}$ & $b$ & $c_{expt}$ & $c$ & $c_{expt}$ &  \\
\hline 
\hline 
\ce{ZrO2}              & $P2_1/c  $     & 5.151    & 5.151~\cite{Howard1988}                        & 5.201             & 5.212                        & 5.323             & 5.317                        &  \\
\ce{HfO2}              & $P2_1/c $      & 5.095    & 5.115~\cite{Adam1959}                           & 5.146             & 5.172                        & 5.267             & 5.294                        &  \\
\ce{WO3}$_m$            & $P2_1/c$       & 7.643     & 7.31~\cite{Loopstra1966}                         & 7.449             & 7.540                        & 7.775               & 7.690                         &  \\
\ce{TiO2}-A     & $I4_1/amd$     & 3.762    & 3.989~\cite{Burdett1987}                        & 3.762             & 3.989                        & 9.603             & 9.522                        &  \\
\ce{TiO2}-R     & $P4_2/mnm$     & 4.581    & 4.593~\cite{Burdett1987}                        & 4.581             & 4.593                        & 2.946             & 2.958                        &  \\
\ce{WO3}$_t $           & $P4_2/mnm $    & 5.238   & 5.25~\cite{Kehl1952}                         & 5.238             & 5.250                         & 3.917             & 3.92                         &  \\
\ce{SnO2}              & $P4_2/mnm $    & 4.742    & 4.737~\cite{Baur1956}                        & 4.742             & 4.737                        & 3.181             & 3.186                        &  \\
\ce{Fe2O3}             & $R\bar{3}c   $     & 5.035    & 5.035~\cite{Cornell2003}                        & 5.035             & 5.035                        & 13.818            & 13.740                        &  \\
\ce{Bi2Se3}            & $R\bar{3}m   $     & 4.139    & 4.134~\cite{NAKAJIMA1963}                        & 4.139             & 4.134                        & 31.124            & 28.634                       &  \\
ZnO               & $P6_3mc  $     & 3.248    & 3.2427~\cite{Sabine1969}                       & 3.248             & 3.2427                       & 5.202             & 5.194                       &  \\
SiC$_h$            & $P6_3mc  $     & 3.071    & 3.067~\cite{Bauer2002}                        & 3.071             & 3.067                        & 10.051            & 10.032                       &  \\
CdSe$_h$           & $P6_3mc $      & 4.353    & 4.30~\cite{madelung2004}                             & 4.353             & 4.300                         & 7.095             & 7.011                        &  \\
CdS$_h$            & $P6_3mc   $    & 4.176   & 4.142~\cite{DALVAND2011}                        & 4.176             & 4.142                        & 6.7849             & 6.724                        &  \\
\ce{Bi2S3}             & $Pnma $       & 3.974   & 3.981~\cite{SHARMA2012}                        & 11.161          & 11.305                       & 11.794            & 11.147                       &  \\
MAPI$_o$           & $Pnma$        & 8.648    & 8.560~\cite{Baikie2013}                         & 9.051             & 8.840                         & 12.953            & 12.580                        &  \\
SiC$_z$            & $F\bar{4}3m  $     & 4.347    & 4.358~\cite{madelung2004}                        & 4.347             & 4.358                        & 4.347             & 4.358                        &  \\
GaP               & $F\bar{4}3m$       & 5.458    & 5.451~\cite{madelung2004}                         & 5.458            & 5.451                        & 5.458             & 5.451                        &  \\
CdSe$_z$           &$ F\bar{4}3m$       & 6.151    & 6.052~\cite{madelung2004}                        & 6.151             & 6.052                        & 6.151             & 6.052                        &  \\
CdS$_z$           &$ F\bar{4}3m $      & 5.895   & 5.818~\cite{madelung2004}                        & 5.895            & 5.818                        & 5.895             & 5.818                        &  \\
STO              & $Pm\bar{3}m  $     & 3.898   & 3.900~\cite{Cao2000}                     & 3.898             & 3.900                          & 3.898             & 3.900                          &  \\
\ce{Cu2O}              & $Pm\bar{3}m $      & 4.288    & 4.270~\cite{Werner1982}                         & 4.288             & 4.270                         & 4.288             & 4.270                        &  \\
\ce{CsPbI3}$_c$         & $Pm\bar{3}m$       & 6.360     & 6.289~\cite{Trots2008}                       & 6.360              & 6.2894                       & 6.360             & 6.2894                       &  \\
Si                & $Fd\bar{3}m$       & 5.433    & 5.430~\cite{madelung2004}                         & 5.433             & 5.430                         & 5.433             & 5.430                         & \\
\hline 
\hline 
\end{tabular}
\label{tbl:structure}
\end{table}

\begin{table}[ht]

\caption{Comparison of the computed (E$_g$, in  eV)  and experimental (E$_g^{\textrm{expt.}}$, in eV) bandgaps. The subscript  $c$, $h$, $o$, $t$, and $z$ stands for cubic, hexagonal, orthorhombic, tetragonal, and zinc-blend-type polymorphs, respectively. \ce{TiO2}-A and \ce{TiO2}-R  indicate the anatase and rutile polytypes, respectively.
}
\begin{tabular}{llll}
\hline 
\hline 
\textbf{Material} & Space Group & E$_g$ & E$_g^{\textrm{expt.}}$\\
\hline 
\hline 
\ce{ZrO2}              & $P2_1/c $      & 5.31                       & 5.30~\cite{Dash2004}                                  \\
\ce{HfO2}              &  $P2_1/c $      & 5.71                       & 5.90~\cite{Ni2008}                                  \\
\ce{WO3}$_m$            & $P2_1/c   $    & 2.63                       & 2.6-3.00~\cite{KOFFYBERG19794,hodes1976,DIQUARTO1981}                                    \\
\ce{TiO2}-A     &$ I4_1/amd$     & 3.76                       & 3.20~\cite{Tang1993}                                  \\
\ce{TiO2}-R      & $P4_2/mnm$     & 3.48                       & 3.03~\cite{Tang1995}                                 \\
\ce{WO3}$_t$            & $P4_2/mnm$     & 3.44                       & 1.80 ~\cite{Iwai1960}                                 \\
\ce{SnO2}              & $P4_2/mnm$     & 3.57                    & 3.60~\cite{Cho2002}                                \\
\ce{Fe2O3}             & $R\bar{3}c  $      & 2.15                       & 2.20~\cite{Mochizuki1977}                                  \\
\ce{Bi2Se3}            & $R\bar{3}m$        & 1.24                       & 0.56~\cite{SHARMA2012}                                 \\
ZnO               & $P6_3mc    $   & 3.45                       & 3.20~\cite{Reynolds1999}          \\
SiC$_h$            & $P6_3mc   $    & 3.17                       & 3.33~\cite{kaplan1995properties}                                 \\
CdSe$_h$           & $P6_3mc $      & 1.54                       & 1.83~\cite{madelung2004}                                    \\
CdS$_h$            & $P6_3mc  $     & 2.19                       & 2.40~\cite{maity2006}                                  \\
\ce{Bi2S3}             & $Pnma  $      & 2.53                       & 1.36~\cite{SHARMA2012}                                 \\
MAPI$_o$           & $Pnma  $      & 2.43                       & 1.51~\cite{Baikie2013}                                 \\
SiC$_z$            & $F\bar{4}3m $      & 2.25                       & 2.45~\cite{madelung2004}                                 \\
GaP               & $F\bar{4}3m   $    & 2.45                       & 2.32~\cite{madelung2004}                                  \\
CdSe$_z$           & $F\bar{4}3m$       & 1.48                        & 1.90~\cite{madelung2004}                                  \\
CdS$_z$            & $F\bar{4}3m$       & 2.11                       & 2.55~\cite{madelung2004}                                 \\
STO               & $Pm\bar{3}m  $     & 3.34                       & 3.25~\cite{vanBenthem2001}     \\
\ce{Cu2O}              & $Pm\bar{3}m $      & 2.00                       & 2.00 ~\cite{CRCHandbook}           \\
\ce{CsPbI3}$_c$         & $Pm\bar{3}m$       & 1.95                       & 1.78~\cite{Sutton2018}                                 \\
Si                & $Fd\bar{3}m$       & 1.27                       & 1.17~\cite{madelung2004}                                 \\
   \hline 
\hline                     
\end{tabular}
\label{tbl:bandgap}
\end{table}

\clearpage
\section{Band structures }
\begin{figure*}[ht]
	\centering
	\includegraphics[width=0.74\textwidth]{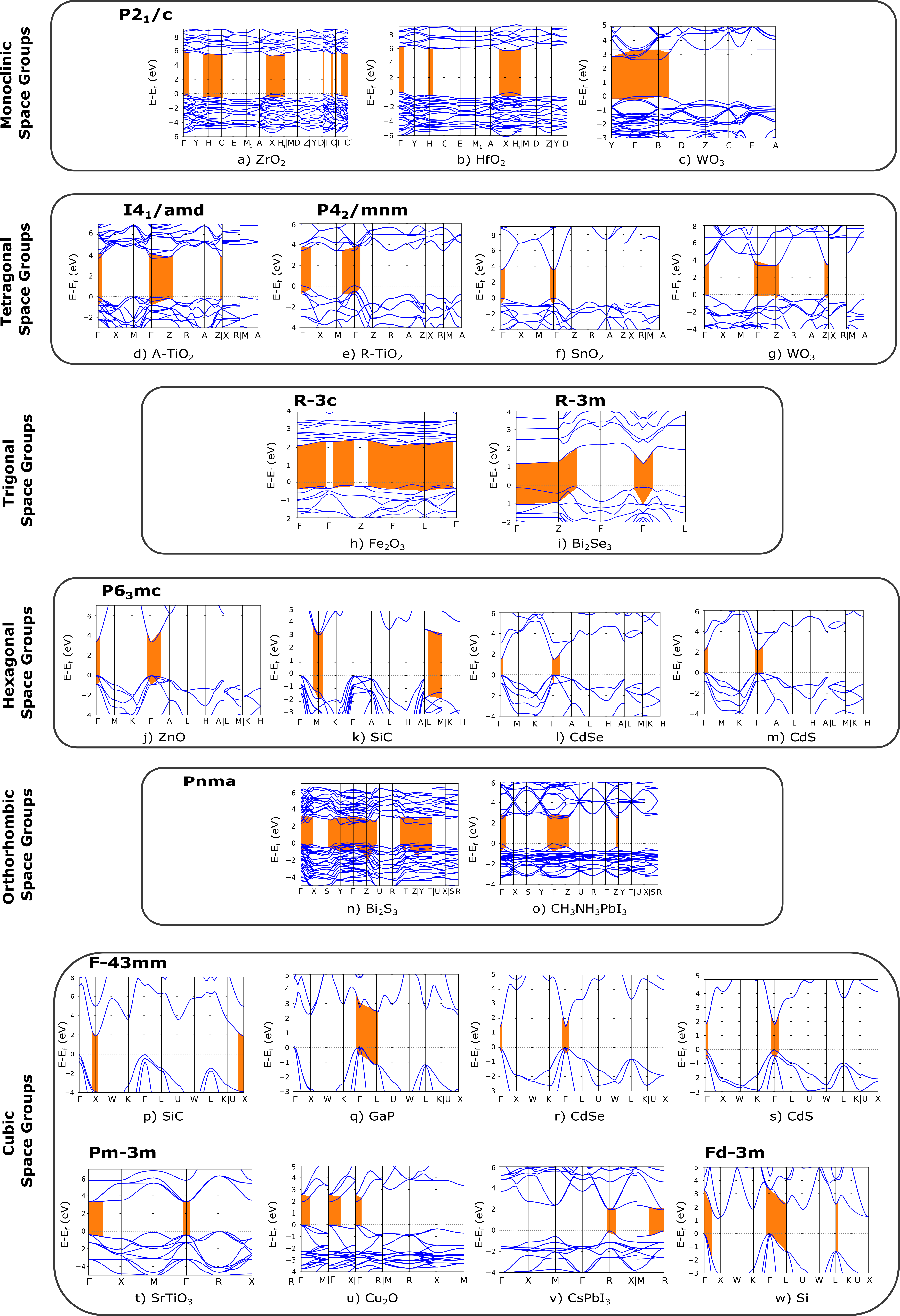}
	\caption{Band structure plots of the materials considered in this work. The shaded regions indicate momentum states where vertical transitions are possible when the photon energy is equal to E$_g$+$\Delta $E (see Table~\ref{tbl:compdetails}).}
	\label{fig:BS}
\end{figure*}

\clearpage
\section{High-symmetry k-points }
\begin{table}[ht]

\caption{High-symmetry k-points for the monoclinic space groups.
}
\begin{tabular}{llllll}
\hline 
\hline 
\textbf{Material} & Space Group & K-point & $\times b_1$ & $\times b_2$ & $\times b_3$ \\
\hline 
\hline 
\ce{ZrO2} & $P2_1/c$ & $\Gamma$ & 0.00 & 0.00 & 0.00 \\
\ce{HfO2} &                & Y & 0.00 & 0.00 & 0.50\\
				&                & H & -0.43 & 0.00& 0.57\\
				&                & C & -0.50 & 0.00& 0.50\\
				&                & E & -0.50 & 0.50& 0.50\\
				&                & M1 & -0.57 & 0.50 & 0.43\\
				&                & A & -0.50 & 0.50 & 0.00\\
				&                & X & -0.50 & 0.00& 0.00\\
				&                & H1 & -0.50 & 0.00& 0.44\\
				&                & M & -0.43 & 0.50 & 0.57\\
				&                & D & 0.00 & 0.50& 0.50\\
				&                & Z & 0.00 & 0.50 & 0.00\\
\hline
\ce{WO3} & $P2_1/c$ & $\Gamma$ & 0.00 & 0.00 & 0.00 \\
				&                & Y & 0.50 & 0.00& 0.00\\
				&                & B & 0.00 & 0.00& 0.50\\
				&                & D & 0.00 & 0.50& 0.50\\
				&                & Z & 0.00 & 0.50& 0.00\\
				&                & C & 0.50 & 0.50 & 0.00\\
				&                & E & -0.50 & 0.50 & 0.50\\
				&                & A & -0.50 & 0.00 & 0.50\\
   \hline 
  \hline                     
\end{tabular}
\label{tbl:monoclinic_skpoints}
\end{table}

\begin{table}[ht]
\caption{High-symmetry k-points for the tetragonal space groups.
}
\begin{tabular}{llllll}
\hline 
\hline 
\textbf{Material} & Space Group & K-point & $\times b_1$ & $\times b_2$ & $\times b_3$ \\
\hline 
\hline 
  \ce{TiO2} & $I4_1/amd$ & $\Gamma$ & 0.00 & 0.00 & 0.00 \\
				&                & X & 0.00 & 0.50 & 0.00\\
				& 					& M & 0.50 & 0.50 & 0.00 \\
				& 					& Z & 0.00 & 0.00 & 0.50\\
				& 					& R & 0.00 & 0.50 & 0.50\\
				& 					& A & 0.50 & 0.50 & 0.50\\
   \hline 
 \ce{TiO2} & $P4_2/mnm$ & $\Gamma$ & 0.00 & 0.00 & 0.00 \\
\ce{SnO2} &                & X & 0.00 & 0.50 & 0.00\\
\ce{WO3}  & 					& M & 0.50 & 0.50 & 0.00 \\
				& 					& Z & 0.00 & 0.00 & 0.50\\
				& 					& R & 0.00 & 0.50 & 0.50\\
				& 					& A & 0.50 & 0.50 & 0.50\\
   \hline
\hline                     
\end{tabular}
\label{tbl:tetragonal_skpoints}
\end{table}

\begin{table}[ht]
\caption{High-symmetry k-points for the trigonal space groups.
}
\begin{tabular}{llllll}
\hline 
\hline 
\textbf{Material} & Space Group & K-point & $\times b_1$ & $\times b_2$ & $\times b_3$ \\
\hline 
\hline 
 \ce{Fe2O3} & $R-3c$ & $\Gamma$ & 0.00 & 0.00 & 0.00 \\
				 &                & L & 0.50 & 0.00 & 0.00 \\
				 &					& Z & 0.50 & 0.50 & 0.50 \\
				 &					& F & 0.50 & 0.50 & 0.00\\		 
   \hline
  \ce{Bi2Se3} & $R-3m$ & $\Gamma$ & 0.00 & 0.00 & 0.00 \\
				 &                & L & 0.50 & 0.00 & 0.00 \\
				 &					& Z & 0.50 & 0.50 & 0.50\\
				 &					& F & 0.50 & 0.50 & 0.00\\
   \hline
\hline                     
\end{tabular}
\label{tbl:trigonal_skpoints}
\end{table}

\begin{table}[ht]

\caption{High-symmetry k-points for the hexagonal space groups.
}
\begin{tabular}{llllll}
\hline 
\hline 
\textbf{Material} & Space Group & K-point & $\times b_1$ & $\times b_2$ & $\times b_3$ \\
\hline 
\hline 
 \ce{ZnO} & $P6_3mc$ & $\Gamma$ & 0.00 & 0.00 & 0.00 \\
\ce{SiC}	 &                & M & 0.50 & 0.00 & 0.00\\
\ce{CdS}	 &                & K & 0.33 & 0.33 & 0.00\\
\ce{CdSe}  &                & A & 0.00 & 0.00 & 0.50\\
				 &                & L & 0.50 & 0.00 & 0.50\\
				 &                & H & 0.33 & 0.33 & 0.50\\
   \hline
\hline                     
\end{tabular}
\label{tbl:hexagonal_skpoints}
\end{table}

 \begin{table}[ht]

\caption{High-symmetry k-points for the cubic space groups.
}
\begin{tabular}{llllll}
\hline 
\hline 
\textbf{Material} & Space Group & K-point & $\times b_1$ & $\times b_2$ & $\times b_3$ \\
\hline 
\hline   
   \ce{Si} & $Fd-3m$ & $\Gamma$ & 0.00 & 0.00 & 0.00 \\
   			 &					& X & 0.50 & 0.00 & 0.50\\
   			 &					& W & 0.50 & 0.25 & 0.75\\
   			 &					& K & 0.375 & 0.375 & 0.75\\
   			 &					& L & 0.50 & 0.50 & 0.50\\
   			 &					& U & 0.625 & 0.25 & 0.625\\
   \hline
   \ce{GaP} & $F-43m$ & $\Gamma$ & 0.00 & 0.00 & 0.00 \\
 \ce{CdS}  			 &					& X & 0.50 & 0.00 & 0.50\\
 \ce{CdSe}  			 &					& W & 0.50 & 0.25 & 0.75\\
 \ce{SiC}  			 &					& K & 0.375 & 0.375 & 0.75\\
   			 &					& L & 0.50 & 0.50 & 0.50\\
   			 &					& U & 0.625 & 0.25 & 0.625\\
   \hline
    \ce{Cu2O} & $Pm3m$ & $\Gamma$ & 0.00 & 0.00 & 0.00 \\
	\ce{SrTiO3}&					& X & 0.00 & 0.50 & 0.00\\
	\ce{CsPbI3}&					& M & 0.50 & 0.50 & 0.00\\
		 			 &					& R & 0.50 & 0.50 & 0.50\\
   \hline
\hline                     
\end{tabular}
\label{tbl:cubic_skpoints}
\end{table}

\begin{table}[ht]

\caption{High-symmetry k-points for the orthorhombic space groups.
}
\begin{tabular}{llllll}
\hline 
\hline 
\textbf{Material} & Space Group & K-point & $\times b_1$ & $\times b_2$ & $\times b_3$ \\
\hline 
\hline 
     \ce{Bi2S3} & $Pnma$ & $\Gamma$ & 0.00 & 0.00 & 0.00 \\
	\ce{CH3NH3PbI3}&					& X & 0.50 & 0.00 & 0.00\\
		 			 &					& S & 0.50 & 0.50 & 0.00\\
		 			 &					& Y & 0.00 & 0.50 & 0.00\\
		 			 &					& Z & 0.00 & 0.00 & 0.50\\
		 			 &					& U & 0.50 & 0.00 & 0.50\\
		 			 &					& R & 0.50 & 0.50 & 0.50\\
		 			 &					& T & 0.00 & 0.50 & 0.50\\
   \hline
\hline                     
\end{tabular}
\label{tbl:ortho_skpoints}
\end{table}

\begin{figure*}[ht]
	\centering
	\includegraphics[width=0.8\textwidth]{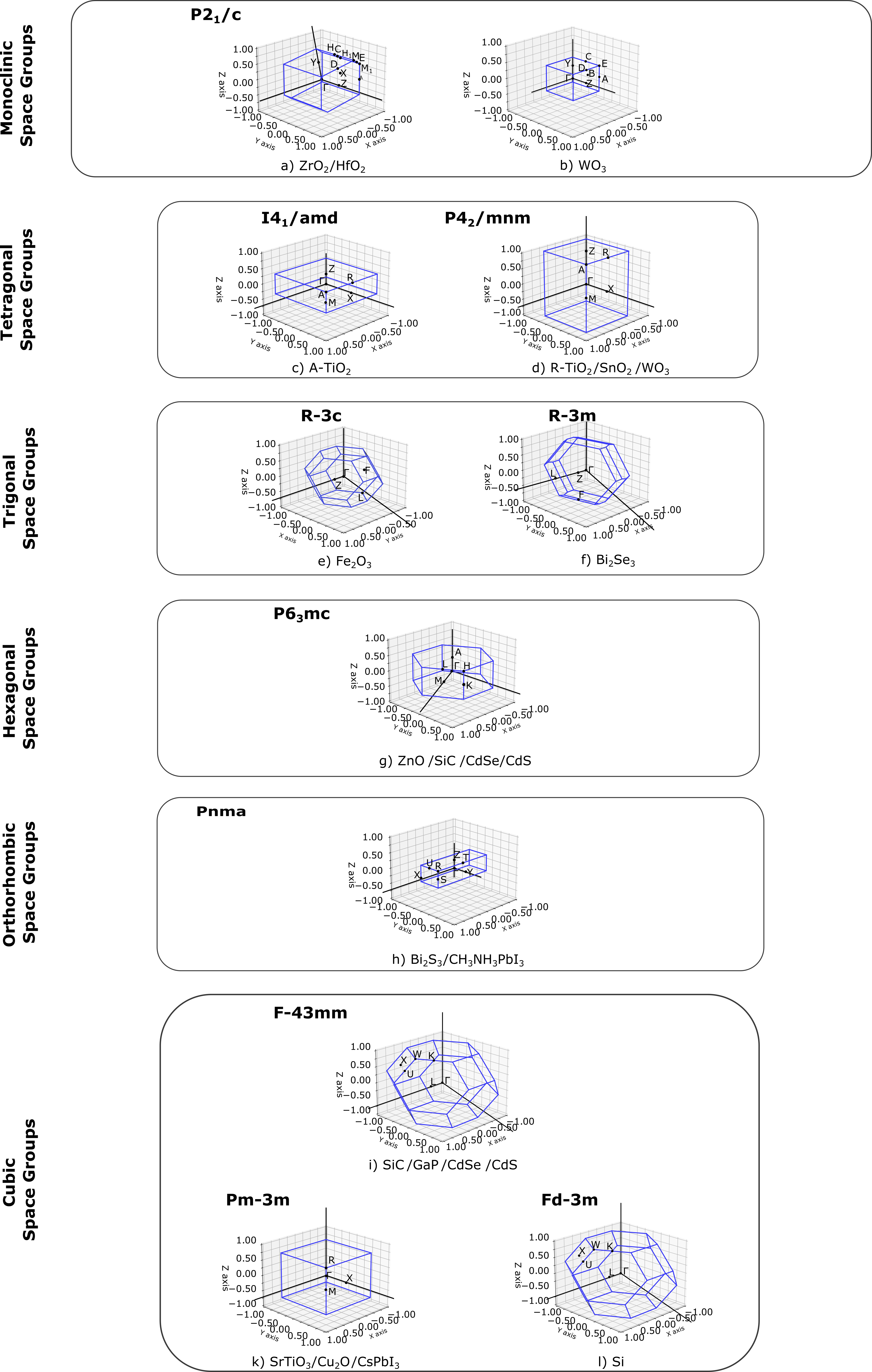}
	\caption{Brillouin zone along with the high symmetry points for different space groups.}
	\label{fig:IBZ}
\end{figure*}
%

\clearpage
\bibliography{references}

\end{document}